\documentclass[12pt,preprint]{aastex}
\usepackage{graphicx} 
\usepackage{natbib}
\newcommand {\hi} {{\rm H}\,{\small\rm I}}

\newcommand {\ovi} {{\rm O}\,{\small\rm VI}}

\newcommand {\siii} {{\rm Si}\,{\small\rm II}}
\newcommand {\siiii} {{\rm Si}\,{\small\rm III}}
\newcommand {\siiv} {{\rm Si}\,{\small\rm IV}}
\newcommand {\cii} {{\rm C}\,{\small\rm II}}
\newcommand {\ciii} {{\rm C}\,{\small\rm III}}
\newcommand {\civ} {{\rm C}\,{\small\rm IV}}
\newcommand {\kms} {\,{\rm km\,s}^{-1}}

\newcommand {\kpc} {\,{\rm kpc}}

\newcommand {\de}{^{\circ}}

\newcommand {\msun}{\,{\rm M}_\odot}

\newcommand {\Gyr}{\,{\rm Gyr}}
\newcommand {\K}{\,{\rm K}}

\newcommand {\msunyr}{\,{{\rm M}_\odot\,\rm yr}^{-1}}
\newcommand {\avg}[1]{\left< #1 \right>} 
\newcommand {\vlos}{v_{\rm LOS}}

\newcommand{\gsim}{\lower.7ex\hbox{$\;\stackrel{\textstyle>}{\sim}\;$}}
\newcommand{\lsim}{\lower.7ex\hbox{$\;\stackrel{\textstyle<}{\sim}\;$}}

\shorttitle{Supernova-driven cooling of the lower Galactic corona}
\shortauthors{Fraternali, Marasco, \& Marinacci}

\begin{document}

\title{Ionized absorbers as evidence for supernova-driven cooling of the lower Galactic corona}

\author{Filippo Fraternali}
\affil{Department of Physics and Astronomy, University of Bologna, via Berti
Pichat 6/2, 40127, Bologna, Italy\\
Kapteyn Astronomical Institute, Postbus 800, 9700 AV, Groningen, NL}
\email{filippo.fraternali@unibo.it}

\author{Antonino Marasco}
\affil{Department of Physics and Astronomy, University of Bologna, via Berti
Pichat 6/2, 40127, Bologna, Italy}

\author{Federico Marinacci}
\affil{Heidelberger Institut f\"{u}r Theoretische Studien, Schloss-Wolfsbrunnenweg 35, 69118 Heidelberg, Germany\\
Zentrum f\"{u}r Astronomie der Universit\"{a}t Heidelberg, Astronomisches Recheninstitut, M\"{o}nchhofstr. 12-14, 69120 Heidelberg, Germany}

\author{James Binney}
\affil{Rudolf Peierls Centre for Theoretical Physics, Keble Road, OX1 3NP, Oxford, UK
}

\begin{abstract}
{
We show that the ultraviolet absorption features, newly discovered in HST spectra, are consistent with being formed in a layer that extends a few kpc above the disk of the Milky Way. 
In this interface between the disk and the Galactic corona, high-metallicity gas ejected from the disk by supernova feedback can mix efficiently with the virial-temperature coronal material.
The mixing process triggers the cooling of the lower corona down to temperatures
encompassing the characteristic range of the observed absorption features, producing a net supernova-driven gas accretion onto the disk at a rate of a few $\msunyr$. 
We speculate that this mechanism explains how the hot-mode of cosmological accretion feeds star formation in galactic disks.
}
\end{abstract}


\section{Introduction}

How the Milky Way and other disk galaxies acquire gas from their surrounding environment is a long-standing problem in galaxy evolution.
Several pieces of evidence, for example studies of the star formation in the Solar neighborhood \citep[e.g.,][]{Aumer&Binney09} and chemical evolution models \citep[e.g.,][]{Chiappini+97}, show that gas accretion is needed to maintain star formation. 
Since their discovery, high-velocity clouds (HVCs) have been regarded as the main source of neutral (\hi) gas accretion in the Galaxy \citep[e.g.,][]{Oort66}. 
However, the recent determination of the distances to the main complexes \citep[e.g.,][]{Wakker+08} has led to estimates of the accretion rate from HVCs that are at least an order of magnitude less than what is required \citep{Putman+12}. 
The same is true for HVCs found in external galaxies \citep{Fraternali09}.
Mergers of gas rich satellites also appear to fall short in providing the required amount of gas accretion \citep{Sancisi+08}.

During the last years, the possibility that gas accretion manifests itself mostly as ionized gas has attracted a growing interest. 
Ultraviolet spectra against bright AGNs obtained with the Space Telescope Imaging Spectrograph (STIS) and the Cosmic Origins Spectrograph (COS) on board of the Hubble Space Telescope (HST) have revealed a number of ionized species at velocities incompatible with those of the Galactic disk material \citep[e.g.,][]{Shull+09}. 
The ions include \cii, \ciii, \civ, \siii, \siiii, and \siiv\ and cover a range of temperatures (assuming collisional ionization equilibrium, CIE) from a few times $10^4\K$ to $2\times10^5\K$. 
They fill about 70-90\% of the sky and they often have neutral gas associated with the absorption \citep{Collins+09,Lehner+12}.
In a few cases they have been identified in the spectra of halo stars, showing that at least a subsample of them lie at distances within about $10 \kpc$ of the Galactic plane \citep{Lehner&Howk11} and ruling out formation in the distant corona or in the Local Group medium.

From a theoretical point of view, the accretion of fresh gas into galaxy disks is a non-trivial problem that has not yet reached a definitive solution. 
The scheme that has recently enjoyed most success is that of the so-called cold flows \citep[e.g.,][]{Dekel&Birnboim06}, where baryons reach the center of a potential well without going through a phase of virialization.
Cosmological simulations suggest that cold flows are the main channel for gas accretion in the early Universe, but they should be replaced by a hot accretion mode for a galaxy like the Milky Way at $z\lesssim1$ \citep[e.g.,][]{Keres+09}.
Thus, the problem of how to cool the hot gas contained in an extended virial-temperature corona and transfer it to the disk to feed star formation still remains. 
The formation of gas clumps in the corona via thermal instabilities has been explored \citep[e.g.,][]{Kaufmann+06}, but it has been ruled out both on theoretical \citep[e.g.,][]{Binney+09,Hobbs+12} and observational \citep[e.g.,][]{Pisano+07} grounds. 
Using hydrodynamical simulations, \citet{Marinacci+10a} showed that gas at the typical temperature of the Galactic corona can cool efficiently if it is mixed with high-metallicity, cooler disk material.
This mixing is enhanced by the onset of a galactic fountain circulation \citep[e.g.][]{Bregman80}.
\citet[][hereafter MFB12]{Marasco+12} used the fountain model of \citet{Fraternali&Binney08}, together with the results of Marinacci et al.'s simulations, to reproduce the kinematics of the \hi\ in the halo of the Milky Way.
In this Letter, we show that the MFB12 model also explains the observational properties of the ionized absorbers detected by \citet{Lehner+12}.

\section{The model}

\begin{figure}
\centering
\setlength\fboxsep{0pt}
\setlength\fboxrule{0.5pt}
\fbox{\includegraphics[width=\textwidth]{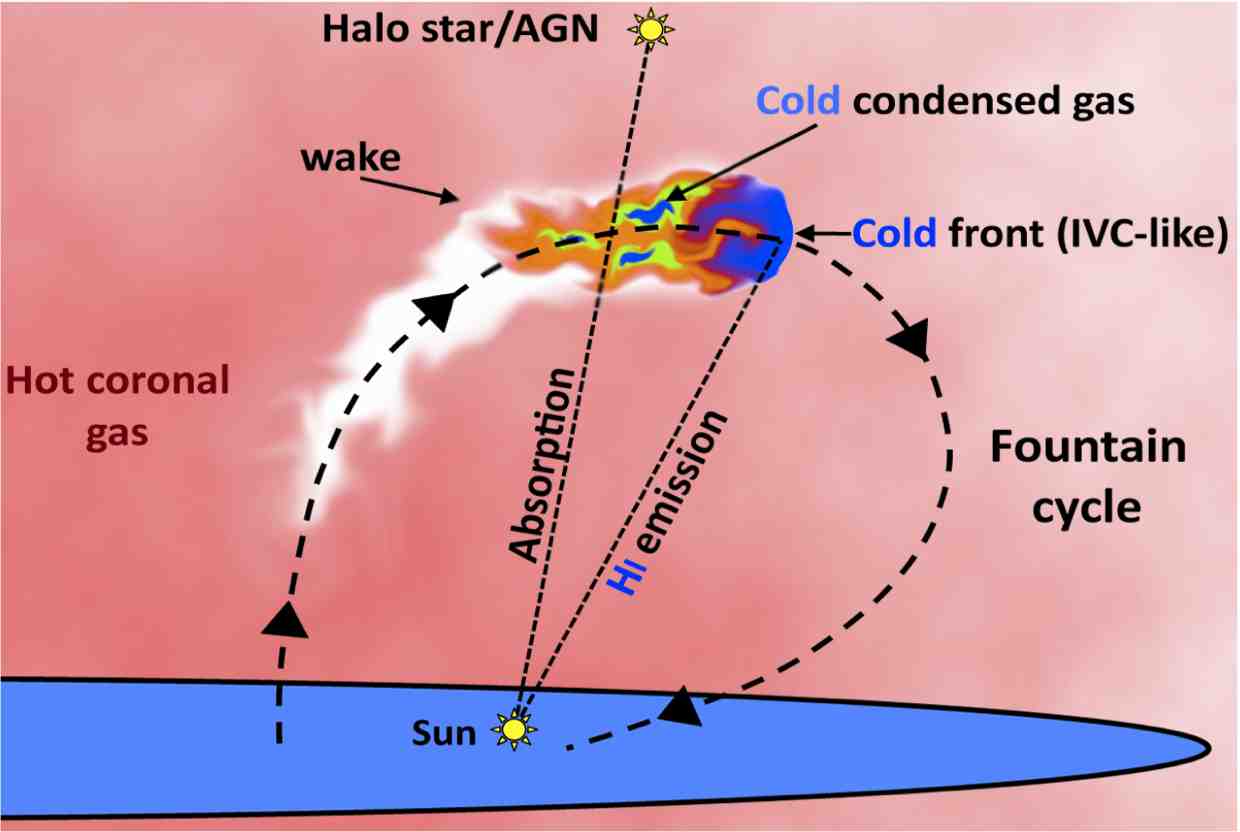}}
\caption{
Fountain clouds, ejected form the disk by supernova feedback, travel through the
halo and interact with the hot gas in the corona.
The interaction creates a turbulent wake where coronal gas is entrained and mixed efficiently with high-metallicity disk material.
This mixing produces a modification of the cold gas kinematics and triggers the cooling of some coronal material, which is then accreted onto the disk when the cloud falls back to it. 
An observer looking towards a background source intercepting the wake detects the absorption lines of the turbulent ionized material. 
An observer looking towards the cold front detects \hi\ emission at velocities typical of intermediate-velocity clouds.} 
\label{scheme}
\end{figure}

We use the supernova-driven accretion model presented in MFB12, where cloud particles are ejected from the disk and move through the halo region interacting with the pre-existing Galactic corona.
The details of this interaction are derived from the hydrodynamical simulations presented in \citet[][this latter hereafter M11]{Marinacci+10a,Marinacci+11}.
In these simulations, a high-metallicity fountain cloud moves through a hot, low-metallicity plasma, representing the Galactic corona. 
Ram-pressure stripping and the onset of the Kelvin-Helmholtz instability produce a turbulent wake in which coronal gas is entrained and mixes efficiently with the disk material. 
The resulting medium at intermediate temperatures and metallicities, trailing the cloud front, can further cool down to recombination temperatures and produce a net transfer of mass from the hot coronal phase to the cold phase.
This condensed coronal material is eventually transferred to the disk where it can feed star formation (see Fig.~\ref{scheme}).
This mass transfer, together with a corresponding exchange of momentum can significantly alter the clouds' trajectories, producing a detectable kinematic signature.
MFB12 used this model to fit the kinematics of the Galactic \hi\ halo in the LAB 21-cm Survey \citep{Kalberla+05} and they determined that the mixing of disk and coronal gas triggers condensation and subsequent accretion of the latter at a rate of $\sim 2 \msunyr$.
Here, we use the same model to predict the properties of the material at intermediate temperatures in the wakes of the fountain clouds and compare them with those of the absorption features detected by \citet{Lehner+12}.
We do not perform a new fit but simply used the best-fit parameters that reproduced the kinematics of the \hi\ halo.

From the simulations of M11 we selected only the gas in the temperature range $4.3\!<\!{\log}(T\!/\!\K)\!<\!5.3$.
This range is representative for the species \siiii, \siiv, \cii, \ciii\ and \civ\, assuming CIE \citep{Sutherland&Dopita93}. 
The gas selected with this temperature cut - hereafter the `warm' gas - is located in the wake of the cloud, typically within 2 kpc of the cold front.
We studied the evolution of this warm gas with time, in particular its mass ratio and velocity difference with respect to the \hi\ (assumed to be at $\log(T\!/\!\K)<4.3$) and its velocity dispersion. 
We found that the warm gas lags $10-20\kms$ behind the cold \hi\ front and develops a turbulent motion with a velocity dispersion of $\sim30\kms$, which dominates over the thermal broadening. 
The details of this analysis are given in a parallel paper (Marasco et al., MNRAS, submitted). 
The dynamical model of MFB12 gives us the column-density of neutral gas as a function of position in the sky and line-of-sight radial velocity.
From this, using the above analysis, we can predict the column-density of the warm gas at every location in the position-velocity space, which can then be compared with the HST absorption data.
The photo-ionizing flux from the disk largely dominates the extragalactic contribution \citep{Shull+09}.
Thus photo-ionization is probably non-negligible in the early stages of the clouds' trajectories: MFB12 found that the best fit to the \hi\ data is obtained if the particles are ionized, and therefore are not visible in \hi, for $30\%$ of the ascending part of their trajectory.
These ionized outflows are included in our model.

\section{Results}

\begin{figure*}
\centering
\includegraphics[width=\textwidth]{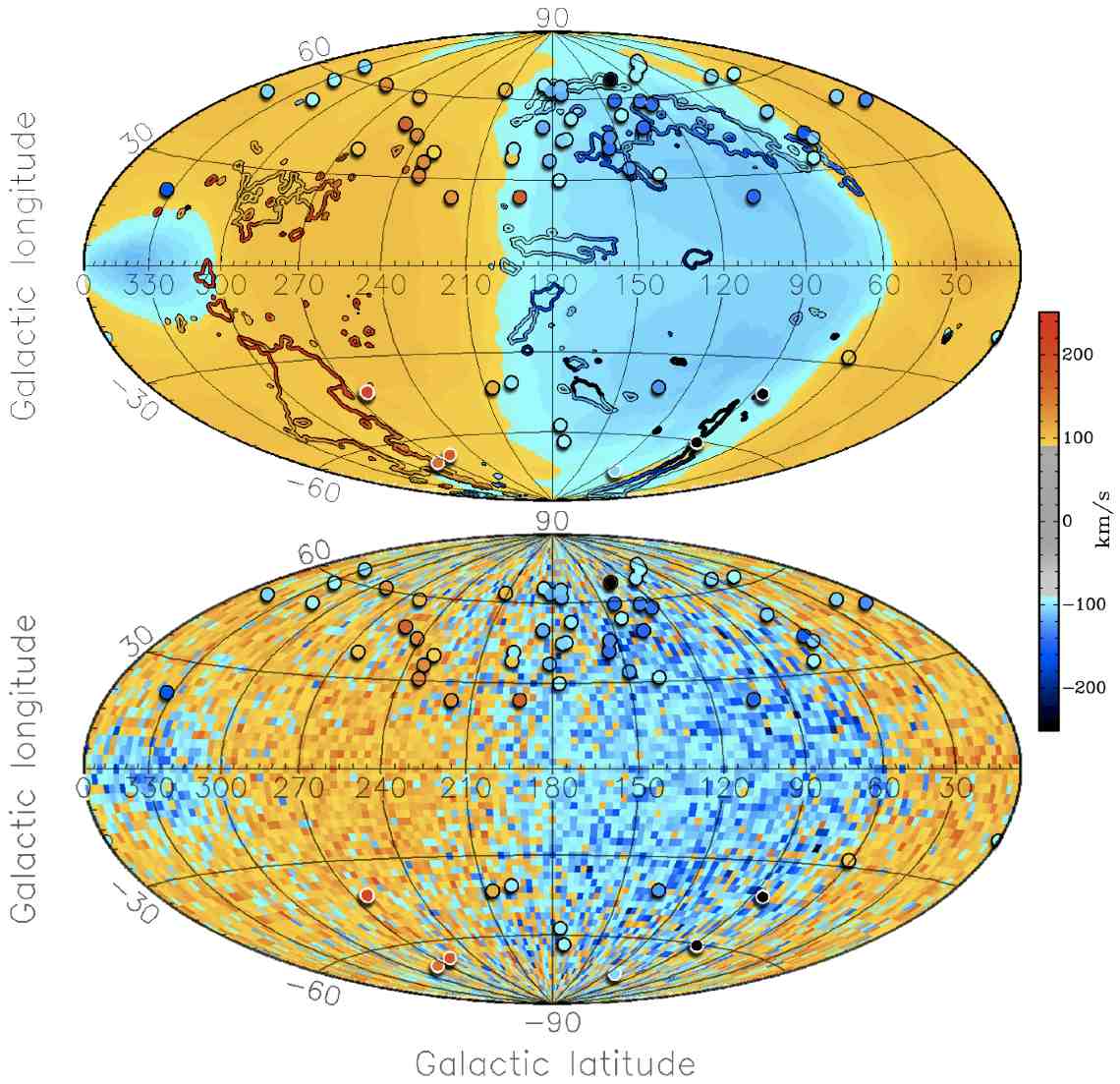}
\caption{All-sky velocity maps comparing the absorption features detected by \citet{Lehner+12} (circles) with the prediction of our model of supernova-driven cooling of the lower corona (background color). 
Multiple detections along the same line of sight are averaged in velocity so the number of points (65) in these panels does not correspond to the total number of detections (84).
Circles with white borders represent absorptions that we consider to be associated with the Magellanic Clouds/Stream. 
The \emph{top panel} shows the \emph{median} velocity field, while the \emph{bottom panel} shows velocities extracted randomly from our model in cells of $2.5\de\times2.5\de$. 
The contours in the top panel show the \hi\ emission from the classical HVCs and the Magellanic Stream, color-coded accordingly to their mean line-of-sight velocities.
}
\label{allsky}
\end{figure*}

Fig.~\ref{allsky} gives an all-sky view of the velocity field predicted by our model compared to the velocity centroids of the absorption lines detected by \citet{Lehner+12}. 
Where there are multiple detections along a single line of sight, we have plotted the average velocity. 
Twelve absorption systems out of 84 are considered to be related to the Magellanic Cloud/Stream and are not taken into account in our analysis.
The top panel of Fig.~\ref{allsky} shows the median velocity field predicted by the model, while in the bottom panel velocities are extracted randomly in cells of $2.5\de\times2.5\de$ to give a measure of the turbulent motions in the wakes.
In both cases, we excluded warm gas at velocities $|\vlos|\!<\!90\kms$, as was done in the observations.
The detections are not distributed isotropically in the sky: the targeted background sources are located mostly at positive latitudes, and no targets are present for $|b|<15\de$. 
Globally, the median velocity field predicts the correct dichotomy between detections at positive and negative velocities, indicating that the absorbing material is consistent with being part of a slowly rotating medium similar to that produced by the interaction between the galactic fountain and the corona
\citep[see also][]{Fox+06,Shull+09}.
However, the data show large fluctuations around the predicted median value. The random velocity field (bottom panel) shows that fluctuations of similar amplitude are present also in our model. 
They are caused by the large velocity dispersion ($\sigma\!=\!30\kms$) of the warm material in the turbulent wakes of the fountain clouds.

\begin{figure}
\centering
\includegraphics[width=0.6\textwidth]{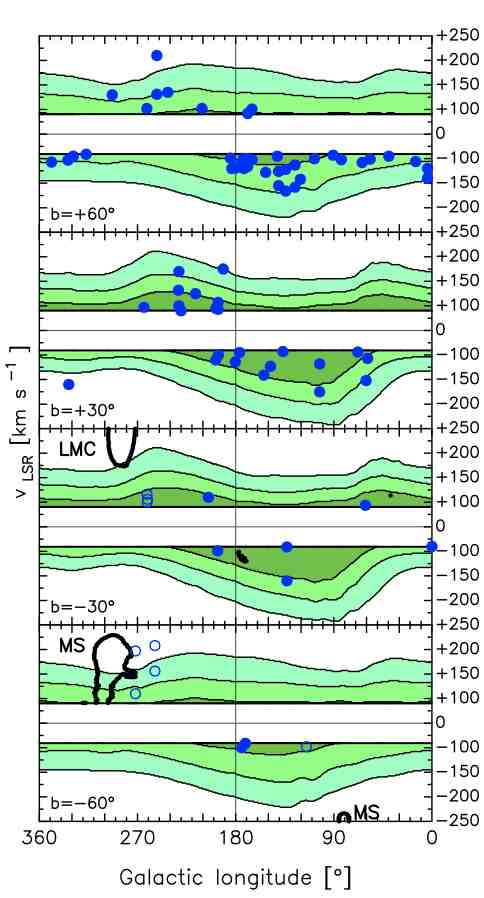}
\caption{
Longitude-velocity diagrams in four different latitude bins (bottom left corners) showing the confidence contours of our model of supernova-driven corona cooling.
The circles show the location in longitude and velocity of the HST absorbers detected by \citet{Lehner+12}.
The empty circles are detections considered related to the Magellanic Clouds/Stream, whose \hi\ emission is shown as the black thick contours.
Five detections at very high-velocities, four of which associated to the Magellanic Stream, are not shown in this plot.
}
\label{lv}
\end{figure}
 
Fig.\,\ref{lv} shows the longitude-velocity distribution of the warm gas in our
model in four different latitude bins. 
The absorption features of \citet{Lehner+12} are over-plotted on these diagrams as filled and empty (if associated to the Magellanic Clouds/Stream) points. 
The contours enclose $68\%$, $95\%$ and $99.7\%$ of the total flux present in the model, and are proportional to the probability of finding a detection at a given position and velocity if the background sources were isotropically distributed in the sky. 
Even though this is not the case, the absorbers follow a trend very similar to that predicted. 
We infer a fraction of detections that is consistent with our model by comparing the predicted and the observed distributions using a Kolmogorov-Smirnov test.
The details are discussed in Marasco et al.\ (MNRAS, submitted), where it is shown that the positions and velocities of $94^{+6}_{-3}\%$ of the UV absorbers are reproduced by our model.
Thus, a large fraction of these ion absorbers are likely to be produced in the wakes of fountain clouds.
These clouds would appear in \hi\ as intermediate-velocity clouds (IVCs) (MFB12) but the ionized gas in the wake may have HVC-like velocities because it has a different kinematics and a much larger turbulence with respect to the \hi.
Interestingly, the fraction of ion absorbers reproduced by our model does not vary significantly if the detections that overlap with the classical \hi\ HVCs are excluded from the calculation.

\citet{Shull+09} obtained an average column density for the high-velocity \siiii\ absorption lines of $\avg{\log N_{\rm Si\,III}}=13.42\pm0.21$, with velocity widths ranging from $40$ to $100\kms$.
We compared this value with the prediction of our model by integrating line profiles of the warm gas over $70\kms$ around the velocity centroid. 
To convert this value into a \siiii\ column density, we further assumed Solar abundance ratios and used the average metallicity of the warm gas in the simulations ($\log[Z/Z_{\odot}]\!=\!-0.24$) with the maximum \siiii\ fraction in CIE \citep[$0.903$ for $\log(T\!/\!K)\!=\!4.45$; see][]{Sutherland&Dopita93}. 
We found a \siiii\ column density, averaged over all the absorbers, of $\avg{\log N_{\rm Si\,III}}=13.44\pm0.36$. 
Hence, the density of gas at intermediate temperatures produced in our model by mixing the fountain material with the hot Galactic corona is in remarkable agreement with the observations.
Note that these column densities are derived assuming CIE, and this agreement may show that the gas is not too far from it.
However, a doubling or tripling of the ion density by photo-ionization would be still compatible within the errors.
We stress again that our model has not been fitted to the absorption data, the only fit that has been performed is on the \hi\ component, which is related to the ionized material only by the cloud-corona interaction and the underlying physics of the turbulent mixing.
Thus, this comparison strongly supports the validity of our approach.

A further comparison between our model and observations comes from counting the number of fountain-cloud wakes intercepted along the lines of sight. 
In our model, the Galactic halo is populated by $\sim10^4$ fountain clouds, given the total \hi\ halo mass ($\sim3\times10^8\msun$; MFB12) and the mass of a typical (intermediate-velocity) cloud \citep[few $\times10^4\msun$;][]{vanWoerden+04}. 
We considered that these $10^4$ clouds are distributed around the Galactic plane with the density profile obtained by MFB12 for the \hi\ halo. 
We assumed that each cloud has an associated wake, for which the volume occupied by the warm gas is derived from the M11 simulations.
%; this can vary by a factor $2\!-\!3$ depending on the initial velocities of the clouds.
With the above information we estimated that an observer placed at the position of the Sun should intercept an average of $0.5-1.0$ wakes per line of sight, which nicely compares with the average number of $\sim0.7$ detections per line of sight in the dataset of \citet{Lehner+12}.

\citet{Sembach+03} and \citet{Savage+03} provide locations, velocities and column-densities for high-ionization (\ovi) absorbers surrounding the Milky Way. 
Marasco et al.\ (MNRAS, submitted) show that more than half of these absorbers are compatible with the same model of supernova-driven accretion presented here. %but there seems to be a second population, which is likely to be located in the outer corona of the Milky Way or around nearby galaxies.
\citet{Shull+09} provide velocity ranges for Si ions (\siii, \siiii, \siiv) found in STIS and FUSE spectra of 37 AGNs.
We used the central velocities of these ranges and found that only 33.4 \% of these features are compatible with our model.
%We used the central velocities of these ranges and found that only 33.4 \% of these features are compatible with our model.
This percentage becomes 63.2 \% when the dataset of \citet{Shull+09} is considered together with that of \citet{Lehner+12}.
%\citet{Lehner&Hock11} provide a sample of STIS absorption features selected with the same criteria adopted by \citet{Lehner+11}.
%Using this sample in combination with \citet{Lehner+11} the reproduced percentage is 70.9 \%.
This result is surprising as it seems to show the two dataset are not fully compatible. 
It appears that most of the difference is due to a number of absorbers at high negative velocities in the region $20<l<120$ in the \citet{Shull+09} sample.
Although this requires further investigation, it is possible that these absorbers lie in a  large ionised envelope surrounding the Magellanic Stream \citep{BlandHawthorn+07}.

\section{Conclusions}

We have shown that the newly discovered UV absorption features in the Galactic halo, observed in HST spectra of bright AGNs and some halo stars, are largely consistent with the predictions of supernova-driven cooling of the Galactic corona.
To make this comparison, we extended the supernova-driven accretion model of MFB12 by including the intermediate-temperature gas generated by the interaction between the galactic fountain clouds and the corona.
We did not perform a new fit but used the best-fit parameters that match the \hi\ kinematics in the LAB survey.
Our model is able to reproduce the positions and velocities of the vast majority of
the detected absorbers. 
Moreover, the model predicts the correct mean column density and the number of intervening absorbers along the line of sight. 

Our findings support the idea that the vast majority of the absorbers are produced in a turbulent multi-phase layer a few kpc thick surrounding the Galactic plane.
This layer is created by supernova feedback, which produces a galactic-fountain cycle that fosters the interaction between the high-metallicity disk material and the corona.
The mixing of the two media lowers the temperature and increases the metallicity of the coronal gas, thus reducing dramatically its cooling time.
As a consequence, part of the lower corona cools to lower and lower temperatures encompassing the range characteristic of the species considered in this work.
When recombination occurs the gas is potentially visible in \hi, but being buried within the fountain cycle it cannot be directly detected. 
This is the reason why cold gas accretion has escaped detection for so long.
The only way to unveil its presence is via the effects that it has on the kinematics of the \hi\ halo \citep[][MFB12]{Fraternali&Binney08}.
This process produces a net accretion of fresh gas from the lower corona onto the disk at a rate of a few $\msunyr$.

\begin{figure}
\centering
\includegraphics[width=\textwidth]{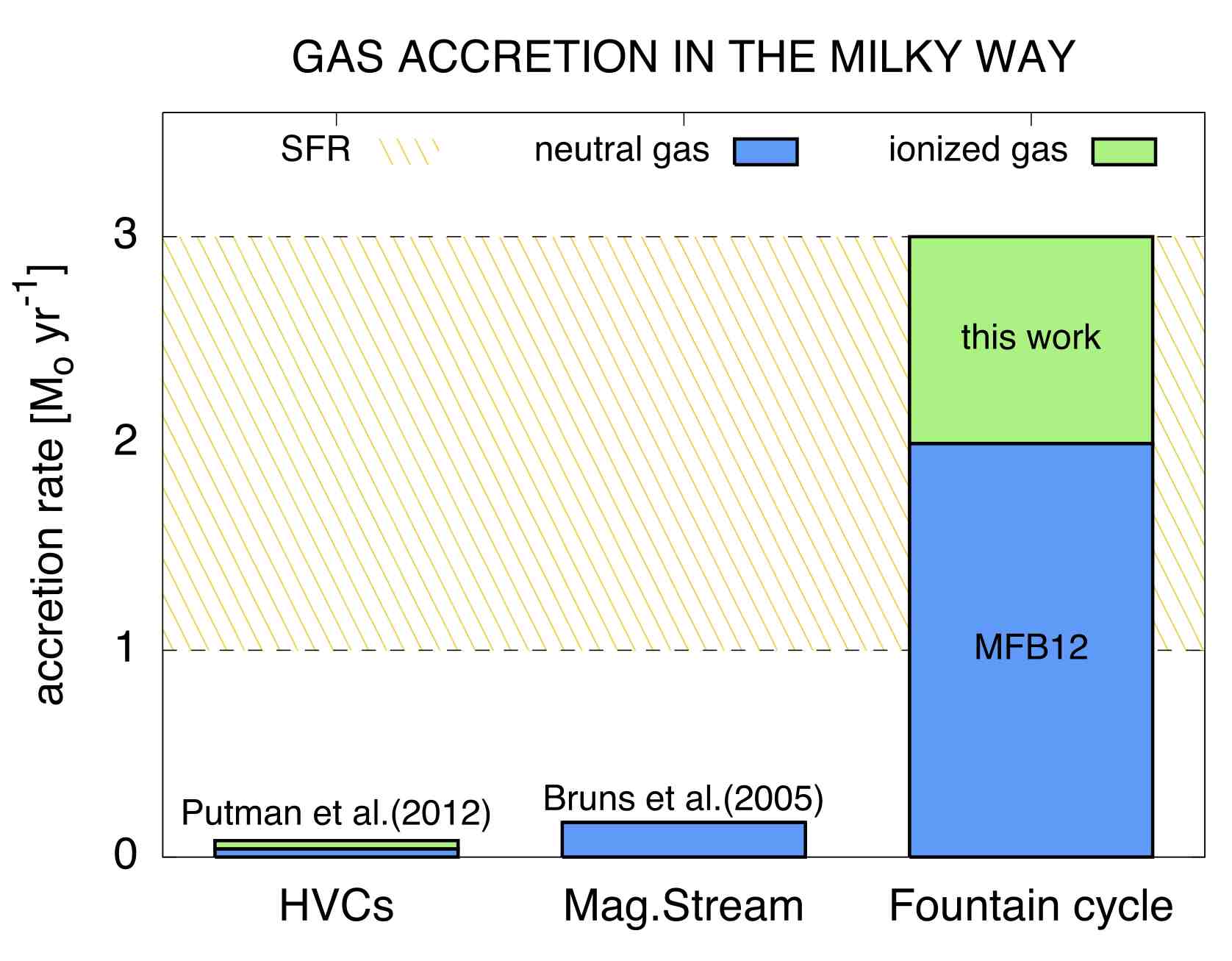}
\caption{
Comparison between different sources of gas accretion for the Milky Way. The estimate for the fountain-cycle (supernova-driven accretion) comes from fitting the kinematics of the \hi\ halo (MFB12) and it is confirmed by the ionized absorption features studied in this work.}
\label{accretionMW}
\end{figure}

In Fig.~\ref{accretionMW} we compare the accretion rates onto the Galaxy that one can infer from the currently available gas sources. 
The classical HVCs contribute only $0.08\msunyr$, and only half of this gas is in the neutral phase \citep{Putman+12}. 
This value is uncertain and somewhat debated but still of the same order of magnitude as the previously often quoted value of $0.1-0.2 \msunyr$ \citep{Wakker+07}.
Direct accretion of cold gas from satellites is only visible in the Magellanic Stream, which is a sporadic event and it is very unlikely to reach the Galactic disk before being ablated and thermalized.
If the Magellanic Stream survives the journey, it will merge with the Galactic ISM on a timescale likely larger than $1\Gyr$ \citep{Putman+12}. 
Given the \hi\ mass of the Stream, $\sim1.2\times10^8\msun$ \citep{Bruns+05}, this gives an upper limit to the accretion rate of $\sim0.16\msunyr$. 
The galactic fountain instead produces an accretion of cold pristine gas onto the disk at a rate of $\sim2\msunyr$ (MFB12).
This value is remarkably similar to the current SFR of the Galaxy, which lies in the range $1-3\msunyr$ \citep[e.g.,][]{Diehl+06,Robitaille&Whitney10}.
In addition, from the present analysis, we infer a further accretion of $\sim1\msunyr$ of gas in the ionized phase, but it is unclear whether this gas can take part in the star formation process or not. 
Note that this value agrees with the estimates of both \citet{Lehner+12} and \citet{Shull+09}.
All the above estimates are corrected for He abundance. 
Clearly, the coronal material harvested by the fountain cycle provides the necessary gas supply for star formation to proceed.

Current cosmological simulations suggest that galaxies above a virial-mass
threshold of a few times $10^{11} \msun$ should acquire gas via hot-mode
accretion, which mainly feeds the hot corona rather than the star formation
in the disk. The galaxy's central black hole accretes gas at a rate that
rises steeply with the corona's central density, and energy released by this
accretion largely offsets the corona's radiative losses, which are dominated
by its dense centre \citep[e.g.,][]{Omma&Binney04}.  The Milky Way and
similar star-forming galaxies became massive enough to enter this regime of
black-hole stabilization at $z\gsim 1$ and consequently, in simulations,
their star formation rates (SFRs) have declined since then by roughly an
order of magnitude.  In contrast, the star formation histories of the
Milky-Way and nearby galaxies of similar masses appear to have declined much
more slowly \citep[e.g.,][]{Panter+07, Fraternali&Tomassetti12}.
Supernova-driven accretion provides an explanation for this apparent
contradiction, as the presence of an active star-forming disk allows the
Galaxy to continuously cool and collect fresh gas from its corona at
significant distance from the centre despite episodic re-heating by
Sgr A*.  
We speculate that supernova-driven gas accretion has been the
way in which the cosmological hot-mode accretion has fed the star formation
in the Milky Way and similar disk galaxies after the initial cold-mode phase.
The ability of the MFB12 model to account so nicely for the HST absorption-line data suggests that this picture is correct.

The same mechanism also explains the dichotomy between the SFRs of blue-cloud and red-sequence galaxies of similar masses.
If star formation were sustained by spontaneous cooling of cosmological coronae, it would be difficult to understand why galaxies residing in similar potential wells - and therefore realistically surrounded by similar coronae \citep{Crain+10} - can have completely different current SFRs.
The puzzle remains even if star formation were fed by the tail of the cold-mode accretion as suggested by \citet{Fernandez+12} among others.
By contrast, in our scheme it all comes down to the presence or not of a (star-forming) disk of cold gas.
The disk effectively acts as a {\it refrigerator} that cools and carves out the corona from below.
As long as a galaxy is able to retain its gaseous disk it can hope to harvest more cold material from the corona, but when the disk is gone it becomes irreversibly ``red and dead''.

\begin{acknowledgments}
{We thank the referee Mike Shull for a constructive report.
FF, AM, and FM thank support from PRIN-MIUR 2008SPTACC. FM also acknowledges support from the collaborative research centre ``The Miky Way system'' (SFB 881) of the DFG.
}
\end{acknowledgments}

\clearpage

%\bibliographystyle{apj}
%\bibliography{myBib}

\end{document}